# Performance of 1-D and 2-D Lattice Boltzmann (LB) in Solution of the Shock Tube Problem

M. Komeili[1], M. Mirzaei[2], M. Shabouei[3]

**Abstract-** In this paper we presented a lattice Boltzmann with square grid for compressible flow problems. Triple level velocity is considered for each cell. Migration step use discrete velocity but continuous parameters are utilized to calculate density, velocity, and energy. So, we called this semi-discrete method. To evaluate the performance of the method the well-known shock tube problem is solved, using 1-D and 2-D version of the lattice Boltzmann method. The results of these versions are compared with each other and with the results of the analytical solution.

## I. INRTODUCTION

IT is about 20 years since Frisch first time (1987) succeeded to use the Lattice Boltzmann Equations to calculate the viscosity of Lattice Gas Cellular Automata (LGCA) [1]. Then, this method has been considered as a new solver of the fluid flow in the various regimes. Ranging from low speed [2-4] to high speed flows [5-8]. In 1987 Alexander introduced a modified LB method to model the compressible flows [9]. The Sung model modifies the LB in such a manner that the fluid velocity is added to particles' velocity [10-17]. Its method regains the results of the Euler and Navier-stockes solutions with first-order and second-order precision, respectively.

In the present work, we intend to develop the Sung method to model 1-D and 2-D compressible flow fields and to evaluate the performance of the model for solution of 1-D problems.

1.  M. Komeili is the Master student of Aerospace Engineering in K.N. Toosi University, Tehran, Iran (e-mail: matin_444@yahoo.com).
2.  M. Mirzaei is Associate professor in K.N.Toosi University, Tehran, Iran (email: mirzaei@kntu.ac.ir).
3.  M. Sabouei is the Master student of Aerospace Engineering in K.N. Toosi University, Tehran, Iran (email:m.shabouei@gmail.com).

## II. LATTICE BOLTZMANN EQUATIONS

The first step in Lattice Boltzmann models is determination of a distribution function. It defines the probability of finding a particle on the specific position and the specific time with a definite velocity.

While, the distribution function is identified throughout the domain. We will be able to calculate the Microscopic quantities (i.e. density, velocity, pressure, temperature, and energy).

The relations between the microscopic quantities and the distribution function are

$$\rho = \int f d\mathrm{v} \qquad (1)$$

$$\rho \mathrm{V} = \int \mathrm{v} f d\mathrm{v} \qquad (2)$$

$$\rho \mathrm{E} = \int e f d\mathrm{v} \qquad (3)$$

Where, is the distribution function, the lower cases relate to particles, and upper cases relate to flow.

So, if we can find the distribution function in each time step our problem has been solved. The distribution function in the next time step is calculated from the following equation:

$$f\left(x + \mathrm{v}\Delta t, \mathrm{t} + \Delta t\right) - f\left(x, \mathrm{v}, t\right) = \Omega\left(x, \mathrm{v}, t\right) \qquad (4)$$

Equation (3) is called BGK equation Where, is the collision operator that can be calculated by:

$$\Omega(x, \mathrm{v}, t) = -\frac{1}{\tau} \left[ f(x, \mathrm{v}, t) - f^{eq}(x, \mathrm{v}, t) \right] \quad (5)$$

$f$ is equilibrium distribution function. This quantity is only a function of microscopic quantities. In the other word, the distribution function and consequently the microscopic values are obtained in the subsequent time if $f$ is identified.

### III. MODIFIED LATTICE BOLTZMANN EQUATION

#### A. Distribution Function

Conventional Lattice Boltzmann models cannot be used for compressible flows due to their limitation in prediction of maximum particles' velocities. In our model in compressible flows, we added flow velocity to particles velocity on each node. So, each particle is translated from a node to any node into the field.

In this case, distribution function depends on the following variables:

$\vec{x}$ ; The position of particle.

$\vec{r}$ ; The migrating velocity which translates particles from the starting points to their destinations.

$\vec{\xi}$ ; The particle velocity.

$\zeta$ ; The particle energy.

t ; The time .

Since is a discontinuous variable, but and are the continuous quantities, we called this method "semi-discrete".

In fact each node carries mass (m), momentum (m $\vec{\xi}$ ), and energy (m $\zeta$ ) to its destination.

The relation between microscopic quantities and the distribution function for this case is as follow:

$$Y = \sum_{\vec{r} \in s} \int_D \eta f(\vec{x}, \vec{r}, \vec{\xi}, \zeta, t) d\vec{\xi} d\zeta \quad (6)$$

Where

$$Y = \left[ \rho, \rho\vec{v}, \rho E \right]^T \quad (7)$$

$$\eta = \left[ m, m\vec{\xi}, m\zeta \right]^T \quad (8)$$

's' is a set which contains all the $\vec{r}$ vectors.

$D = D_1 \times D_2$ which $\vec{\xi} \in D_1$ and $\zeta \in D_2$ .

If we equate relaxation time ( $\tau$ ) to 1 the collision operator converts to:

$$\Omega(x, \mathrm{v}, t) = \left[ f(x, \mathrm{v}, t) - f^{eq}(x, \mathrm{v}, t) \right] \quad (9)$$

So, the well-known BGK will be:

$$f\left(\vec{x} + \vec{r}\Delta t, \vec{r}, \vec{\xi}, \zeta, t + \Delta t\right) = f^{eq}\left(\vec{x}, \vec{r}, \vec{\xi}, \zeta, t\right) \quad (10)$$

#### B. Equilibrium Distribution Function

First, we introduce some quantity in order to define equilibrium distribution function

$$\vec{c}_{jvk} = \vec{v}_k + \vec{c}'_{jv} \quad (11)$$

$$\vec{\vec{c}}_{jvk} = \vec{v} + \vec{c}'_{jv} \quad (12)$$

$$\vec{v}_k = \vec{v} + \vec{v}'_k \quad (13)$$

$$\zeta_{jv} = \frac{1}{2} \vec{c}_{jv}^2 + \phi = \frac{1}{2}\left(v^2 + 2\vec{c}_{jv}.\vec{v} + \left|\vec{c}'_{jv}\right|^2\right) + \phi \quad (14)$$

In the above equations $\left\{ c'_{jv} ; j = 1, 2, ..., b_v \right\}$ is a set which its arrays are vector, $b_v$ defines number of direction on each node and depends on the type of the lattice, index $v$ clarifies number of velocity level on each direction. $\zeta_{jv}$ is any kind of energy except kinetic one and it define by :

$$\phi = \left[ 1 - \frac{D}{2}(\gamma - 1) \right] e \quad (15)$$

D is lattice dimensions; $e$ is interior energy $\left( e = E - \frac{1}{2} v^2 \right)$.

$c'_{jv}$ has symmetric properties. This means that:

$$\sum_{j=1}^{b_v} c'_{jv} = 0 \tag{16}$$

$$\sum_{j=1}^{b_v} c'_{jv} c'_{jv} = \frac{b_v}{D} c'_v I \tag{17}$$

$$\sum_{j=1}^{b_v} c'_{jv} c'_{jv} c'_{jv} = 0 \tag{18}$$

Where, is unit tensor.

Consider $x$ an arbitrary node; $\vec{v}$ is flow velocity on this node; $\vec{v}_k$ is the vector which join $x$ to the nodes that surround; $\vec{v}'_k$ are vectors which connect the end point of $\vec{v}$ to the surrounded nodes (Fig.1).

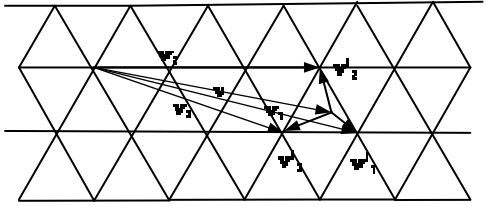

Fig.1. migrating velocity of particles in hexagonal lattice.

Equilibrium distribution function is defined:

$$f^{eq}\left(\vec{x}, \vec{r}, \vec{\xi}, \zeta, t\right) = \begin{cases} d_{vk} \delta\left(\vec{\xi} - \vec{c}_{jv}\right) \delta\left(\zeta - \zeta_{jv}\right) & \vec{r} = \vec{c}_{jvk} \\ 0 & \vec{r} \neq \vec{c}_{jvk} \end{cases} \tag{19}$$

Where $\delta(x)$ has following properties

$$\delta(x) = 0 \quad \text{for} \quad x \neq 0$$

$$\int g(x)\delta(x)dx = g(0)$$

Notice $f^{eq}\left(\vec{x}, \vec{r}, \vec{\xi}, \zeta, t\right)$ is identified on $\left(\vec{r}, \vec{\xi}, \zeta\right) \in S \times D_1 \times D_2$. However, this quantity is nonzero if $\left(\vec{r}, \vec{\xi}, \zeta\right) \in \left\{\vec{c}_{jvk}\right\} \times \left\{\vec{\bar{c}}_{jv}\right\} \times \left\{\zeta_{jv}\right\}$

.

Under these circumstances $f^{eq}_{jvk}$ will be

$$f^{eq}_{jvk}\left(\vec{x}, t\right) = \int_D f^{eq}\left(\vec{x}, \vec{c}_{jvk}, \vec{\xi}, \zeta, t\right) d\vec{\xi}d\zeta = d_{vk} \tag{20}$$

We define $\eta'_{jv}$ as:

$$\eta'_{jv} = \left\{m, m\vec{\bar{c}}_{jv}, m\zeta_{jv}\right\}^T = \left\{m, m\left(\vec{v} + \vec{c}'_{jv}\right), m\left[\frac{1}{2}\left(\vec{v} + \vec{c}'_{jv}\right)^2 + \phi\right]\right\} \tag{21}$$

Substitution of $f^{eq}_{jvk}$ into equation (5), the microscopic quantities are obtained from:

$$Y = \sum_{\vec{r}\in S}\int_D f^{eq}\left(\vec{x}, \vec{r}, \vec{\xi}, \zeta, t\right) d\vec{\xi}d\zeta = \sum_{k,v,j} d_{vk}\eta'_{jv} \tag{22}$$

Where $d_{vk}$ is determined as:

$$d_{vk} = \alpha_k d_v \tag{23}$$

$\alpha_k$ is density fraction and it equals to $\dfrac{\rho_k}{\rho}$

We have chosen triple velocity level for each direction (cell), which are $d_0, d_1,$ and $d_2$. $d_0$ is an arbitrary portion of the total density, we choose a value between 0.4 and 0.55.

$d_1$ and $d_2$ are calculated from the following relations:

$$d_1 = \frac{c'^2_2\left(\rho - b_0 d_0\right) - D(\gamma - 1)\rho e}{b_1\left(c'^2_2 - c'^2_1\right)} \tag{24}$$

$$d_2 = \frac{D(\gamma - 1)\rho e - c'^2_1\left(\rho - b_0 d_0\right)}{b_1\left(c'^2_2 - c'^2_1\right)} \tag{25}$$

$c'_1$ and $c'_2$ are particles module velocity which should be defined such that $d_1$ and $d_2$ have non-negative values.

$$c'_1 = \text{int}\left(\sqrt{\dfrac{D(\gamma - 1)e\rho}{\rho - b_0 d_0}}\right)$$
$$c'_2 = c'_1 + 1 \qquad\qquad\qquad (26)$$

Pressure is obtained from the following equation

$$P = (\gamma - 1)\rho e \qquad (27)$$

### IV. TYPES OF LATTICE

#### A. 1-D Lattice

Fig.2 shows a 1-D lattice. In this lattice, a particle is able to move only from a node to its neighbors located in its right side or left-side.

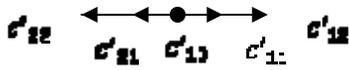

Fig.2.migrating velocity in 1-D lattice

In this figure the migrating velocities are:
$$|c'_{10}| = 0$$
$$|c'_{12}| = |c'_{22}| = 2 \quad , \quad |c'_{11}| = |c'_{21}| = 1$$

Since we have three levels of velocity and two directions, the subscripts are defined as:
$$k = 1 \quad to \quad k = 2$$
$$j = 1 \quad to \quad j = 2$$
$$v = 0 \quad to \quad v = 2$$

Moreover,
$$b_0 = 1 \quad , \quad b_1 = b_2 = 2$$

The contribution of each node to the total density is calculated as:

$$\rho_1 = \rho|v'_2|$$
$$\rho_2 = \rho|v'_1| \qquad\qquad (28)$$

Notice that in this case the velocity has just one direction.

#### B. Square Lattice

If we employ the square lattice (Fig.3) the particles may move in a 2-D space.

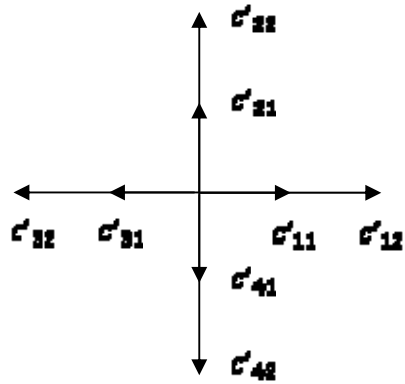

Fig.3.migrating velocity in square lattice

In this figure the migrating velocities $c'_{jv}$ are:

$$|c'_{11}| = |c'_{21}| = |c'_{31}| = |c'_{41}| = 1$$

$$|c'_{12}| = |c'_{22}| = |c'_{32}| = |c'_{42}| = 2$$

Notice that in this case we use dual velocity level and assume $d_0 = 0$. So there are two levels of velocity and four directions:
$$k = 1 \quad to \quad k = 4$$
$$j = 1 \quad to \quad j = 4$$
$$v = 1 \quad to \quad v = 2$$

In addition $b_1 = b_2 = 4$

The contribution of each node which surrounds the end point of vector (Fig. 4) to the total density is defined as:
$$\rho_1 = \rho|u'_3 v'_3|$$
$$\rho_2 = \rho|u'_4 v'_4|$$
$$\rho_3 = \rho|u'_1 v'_1| \qquad\qquad (29)$$
$$\rho_4 = \rho|u'_2 v'_2|$$

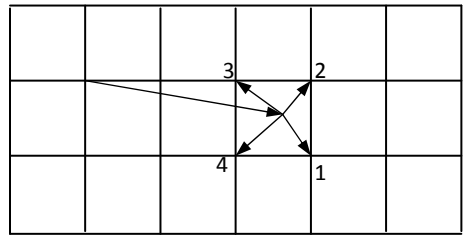

Fig.4.square lattice

## V. NUMERICAL SIMULATIONS

### A. Interior Nodes

Assuming, leads omission of collision step so:

$$f_i^{out}\left(\vec{x}+\vec{c}\,\Delta t, t+\Delta t\right)=f_i^{in}\left(\vec{x}, t\right) \quad (30)$$

First, initial values are set for microscopic flow parameters i.e. Velocity, density, and energy, then calculate translated mass, momentum, and energy from each node (based on velocity level and direction) by the following equations:

$$m_{jvk}=1 \quad (31)$$

$$\xi_{jvk}=\vec{c}_{jv} \quad (32)$$

$$\zeta_{jvk}=\frac{1}{2}\left(v^2+2\vec{x}'_{jv}+\vec{c}'^2\right)+\Phi \quad (33)$$

By multiplying The above quantities by $d_{jv}$ and summation over all $j, v, \text{and } k$ for each node, the microscopic quantities in the next step are obtained.

### B. Boundary Condition

For 1-D lattice, there is no boundary condition. We just add two additional points to both right side and left side of the domain.

For 2-D lattice periodic boundary condition is employed in lateral direction in addition for longitudinal direction two additional columns are use on the limits of the domain.

### C. Shock tube

To evaluate the validity of our models we selected Sod test [12]. The initial conditions are:

$$\rho_L=1.0 \quad ; \quad \rho_R=0.125$$

$$p_L=1.0 \quad ; \quad p_R=0.1$$

$$u_L=1.0 \quad ; \quad u_R=0.125$$

Index L and R show the initial values on the left side and the right side respectively.

A 400 node lattice is adopted for 1-D sock tube problem and a 400×4 node lattice is use for 2-D

shock tube problem. This lattice sizes guarantee the grid independency of the results. The results of 1-D and 2-D solutions were compared with each other and with those of the analytical solution [18] in Fig.5.

### D. Results

The results of the shock tube problem [19,20] are shown in figures 5. Fig 5-a represents the variation of the gas density along the tube. In this figure the results of the 1-D and the 2-D solutions are compared with that of the analytical solution. It can be seen that the 1-D and 2-D solutions leads to the same results and in good agreement with analytical solution. The same trend can be seen for variation if internal energy, velocity and pressure in Figs 5-b to 5-d. notice that the CPU time for 2-D solution is 15 times of that the 1-D solution.

## I. CONCLUSION

We modified the lattice Boltzmann method with adding the flow velocity to the migrating velocities. In this manner the particles can translate to any node in the domain. Accordingly, high Mach numbers flow can be easily simulated. The 1-D lattice is introduced in this paper for the first time. The shock tube problem and other 1-D flows were previously solved using 2-D lattice methods ([14]) but for the first time we develop the 1-D lattice method for such problems. The results of this method are accurate and moreover, its main advantage over 2-D lattice method is its computational efficiency. For instance its CPU time

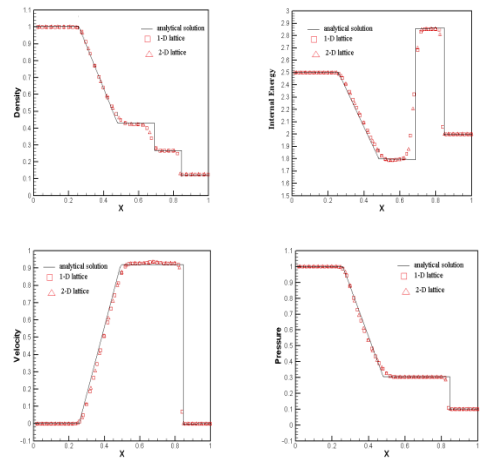

Fig.5. The shock-tube problem: the profiles of density, internal energy, velocity, and pressure at the 75th iteration time. Analytical solution is based on [18]

For shock tube problem was 15 times less than that of the 2-D lattice. The present method may be expanded for more complicated flows (e.g. employing of different gases on the two sides of the tube, consideration of chemical interaction, and reflection of shock on walls).